\documentclass[useAMS,usenatbib]{mn2e}
\usepackage{natbib}
\usepackage{epsfig}
\usepackage{amsbsy}
\usepackage{hyperref}
\providecommand{\adsurl}[1]{\href{#1}{ADS}}

\newcommand{\lya}{Lyman-$\alpha$~}

\newcommand{\be}{\begin{equation}}
\newcommand{\ee}{\end{equation}}
\newcommand{\ba}{\begin{eqnarray}}
\newcommand{\ea}{\end{eqnarray}}
\newcommand{\brr}{\begin{array}}

\newcommand{\err}{\end{array}}
\newcommand{\bc}{\begin{center}}
\newcommand{\ec}{\end{center}}

\def\HeII{\hbox{He$\,\rm \scriptstyle II\ $}}

\DeclareMathAlphabet{\mathsc}{OT1}{cmr}{m}{sc}
\def\testbx{bx}%
\DeclareRobustCommand{\ion}[2]{%
\relax\ifmmode
\ifx\testbx\f@series
{\mathbf{#1\,\mathsc{#2}}}\else
{\mathrm{#1\,\mathsc{#2}}}\fi
\else\textup{#1\,{\mdseries\textsc{#2}}}%
\fi}

\title[Constraints from the \lya forest flux PDF] {Cosmological and
  astrophysical constraints from the \lya forest flux probability
  distribution function}

\author[M. Viel, J.S. Bolton \& M.G. Haehnelt] 
{Matteo Viel$^{1,2}$, James S. Bolton$^{3}$ \& Martin G. Haehnelt$^{4,5}$ \\
$^1$ INAF - Osservatorio Astronomico di Trieste, Via G.B. Tiepolo 11,
I-34131 Trieste, Italy (viel@oats.inaf.it)\\
$^2$ INFN/National Institute for Nuclear Physics, Via Valerio 2,
I-34127 Trieste, Italy\\
$^3$ Max-Planck Institut f{\"u}r Astrophysik,
Karl-Schwarzschild Str. 1, 85748 Garching, Germany\\
$^4$ Institute of Astronomy, Madingley Road, Cambridge, CB3 0HA\\
$^5$ KICC-Kavli Institute of Cosmology, Cambridge \\}

\begin{document}
\maketitle
\begin{abstract}
We use the probability distribution function (PDF) of the \lya forest
flux at $z=2-3$, measured from high-resolution UVES/VLT data, and
hydrodynamical simulations to obtain constraints on cosmological
parameters and the thermal state of the intergalactic medium (IGM) at
$z\sim 2-3$.  The observed flux PDF at $z=3$ alone results in
constraints on cosmological parameters in good agreement with those
obtained from the WMAP data, albeit with about a factor two larger
errors.  The observed flux PDF is best fit with simulations with a
matter fluctuation amplitude of $\sigma_8=0.8-0.85 \pm 0.07$ and an
inverted IGM temperature-density relation ($\gamma\sim 0.5-0.75$),
consistent with our previous results obtained using a simpler
analysis. These results appear to be robust to uncertainties in the
quasar (QSO) continuum placement.  We further discuss constraints
obtained by a combined analysis of the high-resolution flux PDF and
the power spectrum measured from the Sloan Digital Sky Survey (SDSS)
\lya forest data.  The joint analysis confirms the suggestion of an
inverted temperature-density relation, but prefers somewhat higher
values ($\sigma_8 \sim 0.9$) of the matter fluctuation amplitude than the
WMAP data and the best fit to the flux PDF alone. The joint analysis of
the flux PDF and power spectrum (as well as an analysis of the power
spectrum data alone) prefers rather large values for the temperature
of the IGM, perhaps suggesting that we have identified a not yet
accounted for systematic error in the SDSS flux power spectrum data or
that the standard model describing the thermal state of the IGM at
$z\sim 2-3$ is incomplete.  \end{abstract}


\section{Introduction}

The \lya forest is an important cosmological observable that probes
matter density fluctuations in the IGM over a unique range of
redshifts, scales and environments. Many attempts have been made to
measure physical properties of the IGM using \lya forest data. The two
most common approaches are either based on decomposing the information
encoded in the transmitted flux via Voigt profile fitting or treating
the flux as a continuous field with directly measurable statistical
properties (e.g.
\citealt{rauch97,rauch98,theuns98,croft02,meiksin07}). In the second
approach, measurement of the zero, one, two-point or three-point
probability distribution functions (i.e. the mean flux level, the flux
PDF, the flux power and bispectrum) enable a variety of physical
properties to be explored. The mean flux level for example, is
sensitive to the amplitude of the meta-galactic UV background
(\citealt{tytler04,bolt05}) while the flux PDF is sensitive to the
thermal evolution of the IGM \citep{mcquinn09,bolton09}. The flux
power spectrum has been used to constrain cosmological parameters and
the behaviour of dark matter at small scales
\citep{vhs,seljak06,viel08} and the flux bispectrum can be used to
search for signatures of non-gaussianities in the matter distribution
\citep{viel09}.  Ideally, a given IGM model described by a set of
cosmological and astrophysical parameters should agree with all these
statistics  including the  results from Voigt profile decomposition 
at the same time.  In practice, the interpretation of the data is
complex and is heavily dependent on numerical simulations that
incorporate the relevant physical ingredients, but have a limited
dynamic range.

The data  used for these investigations consist mainly of two
kinds of sets
of QSO spectra: the SDSS low--resolution, low signal--to--noise sample
and  UVES/VLT  or HIRES/KECK samples of high--resolution spectra.  The 
characteristics of the low and high-resolution  data sets  are very 
different (the number of SDSS spectra
is about a factor $\sim 200$ larger than that of high-resolution
samples,  but the latter probes smaller scales due to the higher
spectral resolution). Measurements based on \lya forest data have
reached a level of accuracy where an understanding of systematic
uncertainties at the percent level or below (the magnitude of statistical errors
associated with the SDSS sample) has become important.  In this
Letter, we will revisit the flux PDF, which has been investigated previously by several authors
either on its own
(e.g. \citealt{mcdonald00,jena,Becker:2006qj,lidz06,tkim,Bolton08}),
or jointly with the flux power spectrum
\citep{meiksin01,zaroubi06,desjacques07}.  
We shall focus on the flux PDF of the UVES/VLT data as 
recently measured by \cite{tkim} (hereafter K07); the systematic and
statistical errors for this sample have been addressed in
detail. Recently, both \cite{Bolton08} (from the K07 UVES/VLT data)
and \cite{Becker:2006qj} (from independent HIRES/KECK spectra) have found
evidence for a  density-temperature relation of the IGM which appears 
inverted if approximated as a power--law ($\gamma<1$ for
$T=T_0(1+\delta)^{\gamma-1}$).  Here, we will
improve on the analysis performed in \cite{Bolton08} (hereafter B08)
and check its robustness by fully exploring the cosmological and
astrophysical parameter space.  In addition, we briefly discuss a
joint analysis of the flux PDF and SDSS flux power spectrum, and the
possible implications for constraints on cosmological parameters
describing the linear matter power spectrum and the thermal history of
the IGM.

\section{Method}

We use simulations performed with the parallel hydrodynamical
(TreeSPH) code {\small {GADGET-2}} \citep{springel} to calculate the
flux statistics for models with a wide range of cosmological and
astrophysical parameters by expanding around a reference model.  For
the reference model we choose here the 20-256 simulation of B08.  We
refer the reader to this paper for further details, including
resolution and box size convergence tests (see \cite{bb09} for recent
convergence tests on SPH simulations).  We will compare these simulations to
improved measurements of the PDF made by K07 in three redshift bins at
$\langle z \rangle =2.07$, $\langle z \rangle =2.52$ and $\langle z
\rangle =2.94$ based on a set of 18 high resolution ($R \sim 45~000$),
high signal--to--noise (S/N $\geq 30-50$) VLT/UVES spectra.  Further
details regarding the observational data and its reduction, with
particular emphasis on metal removal and continuum fitting errors, may
be found in K07.  In all instances the mock QSO spectra have been
processed to have the same instrumental properties as the observed
data: i.e. the same signal--to--noise, resolution and pixel size.

We explore the following cosmological and astrophysical parameters:
$\Omega_{\rm m}$, $n_{\rm s}$, $H_0$, $\sigma_8$ for the cosmological
part and $T_0^{A,S}(z=3)$ and $\gamma^{A,S}(z=3)$ for the IGM thermal
history, where $A$ and $S$ indicate the amplitude and slope for the
temperature and $\gamma$ relations normalised at $z=3$
($y=A[(1+z)/4]^S$). The amplitude and slope of the effective optical
depth evolution, $\tau_{\rm eff}=-\langle F \rangle$, are varied
assuming a power--law evolution with redshift in order to
conservatively span the observed range suggested by high-resolution
and low-resolution data sets.  We furthermore varied the reionization
redshift ($z_{\rm re}=9$ in our reference model) but found this had no
impact on the flux PDF at $z<3$ (although the differences in ``Jeans
smoothing'' will be important at redshifts close to $z_{\rm re}$,
e.g. \cite{Pawlik09}). For the effect on the flux power we refer to
\cite{mcdonald05} and \cite{vielhaehnelt06}. We also consider the
effect of a misplaced continuum level by adding an extra parameter
$f_{\rm c}$ (the flux following a continuum correction is assumed to
be $F\times (1+f_{\rm c})$).

We compute derivatives of the flux statistics from the 20-256 model at
second order using between two and four simulations for each
cosmological and astrophysical parameter.  For the thermal history we
explore a wide range of possible $T_0$ and $\gamma$ values by
extending the original grid of simulations presented in B08.  The
20-256 model has $\gamma \sim 1.3$ below $z=3$ and the temperature at
mean density in the three PDF redshift bins are
$T_0=14.8,17.6,20.8\,\times 10^3$ K.  This model was shown to be a
poor fit to the K07 data in the simple analysis performed by B08.  Here
we will calculate the $\chi^2$ of our models varying {\it all} the
parameters that affect the flux PDF and not just the effective optical
depth, enabling us to expand around this model.  This simple
Taylor--expansion method was introduced in \cite{vielhaehnelt06} in
order to explore constraints for the SDSS flux power spectrum.  It has
the advantage of enabling the exploration of the parameter space close
to the reference model with an accurate set of hydrodynamical
simulations.  However, the full parameter space cannot be probed
in this way with the same high accuracy (see \citealt{mcdonald05} for
a different approach).

\section{Results for the flux PDF}

\begin{table}
\begin{tabular}{llll}
\hline
{\tiny Param.} & {\tiny (20,256)$^{0.1-0.8}$} &  {\tiny (20,256)$^{0-0.9}$} & {\tiny (20,256)$^{0-1}$} \\
\hline
$\sigma_8$ & $0.81 \pm 0.07$ $(0.80)$ &$0.85 \pm 0.07$& $0.86 \pm 0.06$\\ 
$n_s$ & $0.96 \pm 0.03$ $(0.95)$ & $0.96 \pm 0.03$ & $0.96 \pm 0.03$ \\
$\Omega_{\rm 0m}$ & $0.23 \pm 0.07$  $(0.19)$ & $0.22 \pm 0.06$ & $0.22\pm 0.06$ \\
$H_0$ & $82 \pm 7$ $(80)$ & $84\pm 6$ & $85 \pm 8$ \\
$T_0$ & $19 \pm 6$ $(15)$ & $24 \pm 8$ & $26\pm 7$ \\
$T_0^{s1}$ & $0.6 \pm 1.4$ $(-0.6)$ & $1.3\pm1.2$ & $1.5 \pm 1.0$ \\
$\gamma^A$ & $0.75 \pm 0.21$  $(0.72)$  & $0.51 \pm 0.13$ & $0.51 \pm 0.13$ \\
$\gamma^{s1}$ & $-1.0 \pm 1.0$ $(-1.7)$ & $-1.6\pm 0.9$ & $-1.3 \pm 1.0$ \\
$\tau_{\rm eff}^A$  & $0.312 \pm 0.012$  $(0.312)$ & $0.321\pm 0.010$ & $0.324 \pm 0.010$\\
$\tau_{\rm eff}^{s1}$ & $3.17 \pm 0.18$ $(3.20)$ &$3.16 \pm 0.13$ & $3.25 \pm 0.14$ \\
$f_{\rm c}\times 100$ & $0 \pm 1$  $(0)$ & $-0.8\pm0.4$ & $-1 \pm 0.4$ \\
\hline
$\chi^2/d.o.f.$ &  $35.2/36 $  &  $45.6/48 $ & $64/54 $ \\
\end{tabular}
\caption{Marginalised cosmological and
  astrophysical parameters derived from fitting the flux PDF 
  at $z=2.07,2.52,2.94$ in the flux ranges F=[0.1-0.8],
  F=[0.1-0.9] and F=[0-1] (left, middle and right columns
  respectively): $T_0$ is measured in units of $10^3$ K, $H_0$ in
  km/s/Mpc.   The probabilities of having a
  $\chi^2$ larger than the obtained values are 50, 57 and 17\%,
  respectively.  The values in parentheses are best-fit values 
   for (20,256)$^{[0.1-0.8]}$.}
\label{tab1_pdf}
\end{table}

\begin{figure}
\begin{center}
\includegraphics[width=8cm]{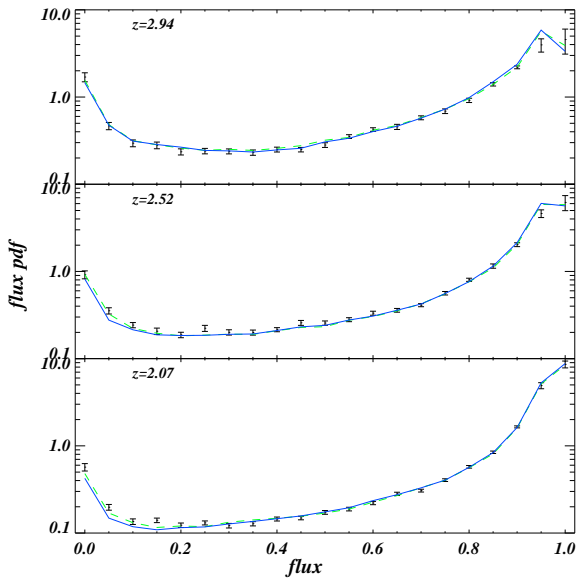}
\end{center}
\caption{Best fit to the flux PDF for model (20-256)$^{0.1-0.8}$
  (continuous blue) and (20-256)$^{0-1}$ (dashed green) in the three
  redshift bins at $\langle z \rangle =(2.07,2.52,2.94)$.}  
\label{fig_bestfitpdf}
\end{figure}

We obtain the best fit to the observed flux PDF for three 
flux intervals $F=[0.1-0.8]$, $F=[0.1-0.9]$, $F=[0-1]$. 
Different flux levels are subject to different
systematic effects, such as the presence of noise and strong
absorption systems at $F\sim 0$ and the effect of continuum fitting
errors at $F\sim 1$ (see K07 for details). Since the PDF error bars
are correlated we expect these systematic errors to nevertheless
impact on the  PDF over the full flux range. The level of 
consistency between the fits to these three 
flux intervals should indicate to what extent these systematic errors 
may or may  not affect the results.

\begin{figure}
\begin{center}
\includegraphics[width=8.5cm,height=10cm]{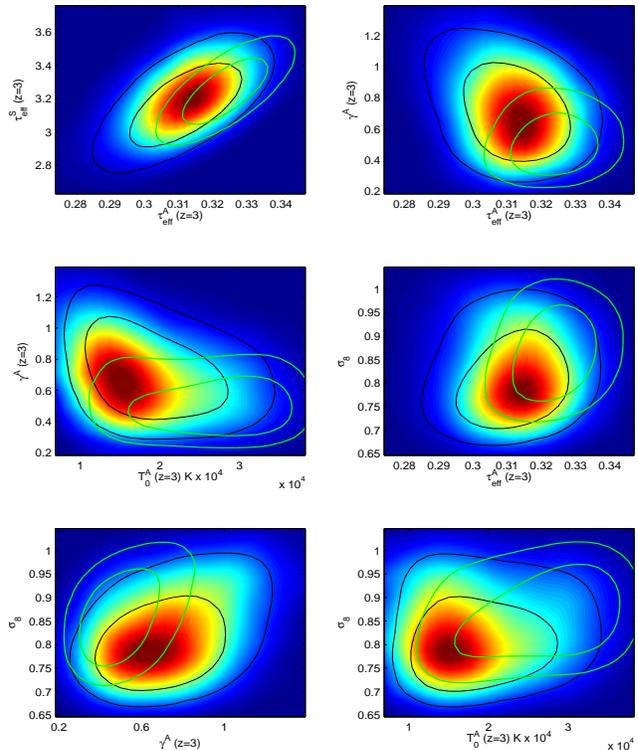}
\end{center}
\caption{2D marginalized and mean likelihood contours for
  cosmological and astrophysical parameters for flux PDF model
  (20,256)$^{0.1-0.8}$ (continuous curves and filled contours). Marginalized
  likelihoods for the model (20,256)$^{0-1}$ are also shown in green.}
\label{fig_2Dpdf}
\end{figure}

For the flux range $F=[0.1-0.8]$ we have a total of 45 data points to
fit and a set of 9 free parameters that will be varied in the Markov
Chain Monte Carlo routines.  We use the following priors on the
effective optical depth, $\tau_{\rm eff}^A=0.36\pm 0.11$ and
$\tau_{\rm eff}^S=3.65\pm 0.21$, based on the observational results
obtained by K07. Note, however, that the final results are affected
very little by the choice of these priors; the constraints on the
effective optical depth amplitude at $z=3$ are in fact much tighter
than these priors assume.  The results of this analysis are summarised
in Table \ref{tab1_pdf} and Figs. \ref{fig_bestfitpdf} and
\ref{fig_2Dpdf}. We obtain a very good fit to the flux PDF for the
flux range $F=[0.1-0.8]$ (reduced $\chi^2=0.98$, a $\chi^2$ larger
than this has 50 per cent probability).  With the data points at
$F<0.1$ we obtain slightly larger values for $\sigma_8$ and $T_0$, but
the results are in agreement at the $1\sigma$ level.  Adding  the
flux range at $F>0.9$ results in a poor fit unless the error bars
on the last two data points are increased by a factor of four.  Note that the covariance properties of these data points are 
strongly  influenced by the choice of the continuum level.  Increasing the
error bars by this factor would account for a misplacement of the
continuum level by a few percent. We also find evidence ($2\sigma$)
that the data prefer a non-zero continuum offset, $f_{\rm c}$, when we
add the regions at low and high transmissivity, but note that realistic errors
in the assumed continuum level should depend on the flux and noise
level and are expected to vary along the spectrum.

In order to further explore the sensitivity to the last two data points
($F=0.95,1$) in each redshift bin, we performed an additional analysis
by combining the two data points for the highest flux levels  into one 
point at $F=0.975$. We then recomputed
the data covariance matrix and the PDF derivatives (without
multiplying the covariance values of this data point by four).  In
this instance the results are  consistent with
those for the $F=[0-0.9]$ and $F=[0-1]$ flux ranges range
to within $1\sigma$. In this case we obtain
$\chi^2=60.6$ for 51 d.o.f., which indicates  a reasonable fit 
(the probability for a value larger than this  is 17\% ). We therefore 
conclude from the results in Table \ref{tab1_pdf}  that our findings
are robust to  continuum fitting uncertainties and that the impact 
of continuum uncertainties is mainly restricted to  the flux range 
$F=0.975-1.025$.

The effective optical depth is constrained very well by the data with
a best-fit value of $\tau_{\rm eff}(z=3)=0.31\pm0.01$, consistent with
observational measurements from high-resolution spectra.  The
constraints on the temperature density relation at $z=3$ are
$(T_0^A,\gamma^A)=(19\pm 6\,,0.75\pm0.21)$, in agreement with the
findings of B08, and no significant evolution in the equation of state
below $z<3$ is inferred (in agreement with \cite{schaye00,ricotti00}).
The PDF alone provides interesting constraints on cosmological
parameters describing the evolution of the linear power spectrum;
$\sigma_8$ and $n_{\rm s}$ are constrained to be in the range
$\sigma_8=0.8-0.85 \pm 0.07$ and $n_{\rm s}=0.96 \pm 0.03$. The derived
cosmological parameters are in good agreement with the results of
other large-scale structure probes such as WMAP and weak lensing data,
albeit with about a factor two larger errors than those from the WMAP
data (e.g. \cite{lesg}).

Our results corroborate the suggestion of an inverted
temperature-density relation at $z=3$.  As for the analysis of the
full flux range the statistical significance of the data favouring an
inverted temperature-density relation $\gamma < 1$ is about
$3\sigma$ at $z \sim 3$. At $z<3$ the data is consistent with an
isothermal ($\gamma \sim 1$) temperature-density relation The results
regarding the thermal state of the IGM do not change significantly if
we omit the  flux range $F >0.9 $.  If we discard both the flux
ranges at low and high emissivity and consider only the flux range
$F=[0.1-0.8]$, there is still evidence for an inverted $T-\rho$
relation, but at a reduced level of significance ($1-1.5\sigma$
C.L.).  The likelihood contours in Fig.~\ref{fig_2Dpdf} 
indicate that a value of $\gamma \sim 1.3$
suggested recently by the \HeII reionization simulations of
\cite{mcquinn09} is between $2$ and $3-3.5\sigma$ discrepant with
the marginalized value we obtain when fitting the flux
range $F=[0.1-0.8]$ and the full flux range, respectively.

\section{Adding the flux power spectrum}

In this section we first revisit constraints from the SDSS flux power
spectrum (PS) alone before proceeding to combine this data set with
the flux PDF of the UVES/VLT data for a joint analysis.  The SDSS flux power spectrum is
based on $3035$ QSO spectra with low resolution and low
signal-to-noise, spanning the redshift range $z=2.2-4.2$ (measurements
are made at 11 wavenumbers in 12 redshift bins). Dealing with the
systematic uncertainties of low resolution and low signal--to--noise
QSO spectra and extracting the flux power is a difficult task. We
refer to \cite{mcdonald05} for a comprehensive study of the removal of
continuum fluctuations, metal line contamination, damped \lya systems
and dealing with the resolution of the spectrograph and noise level in
each of the redshift bins.  All these effects need to be properly
taken into account, as a poor treatment would impact the obtained flux
power in a non-trivial way. In the following, we will use the flux
power provided by the SDSS collaboration, introducing ``nuisance
parameters'' for the resolution and noise in each redshift bin as
suggested by \cite{mcdonald05}, and implicitly assuming that all the
contaminants above have been either removed or properly modelled.

We compute the constraints from the SDSS flux power spectrum in a
similar way to \cite{vielhaehnelt06} with the notable difference that
we calculate the predicted flux statistics by expanding around a model
with $\gamma \sim 1$ in the redshift range $z=[2-4]$, while the
original analysis was based on simulations with $\gamma\sim 1.6$.
Furthermore, the flux power is computed using a Taylor expansion to
second instead of first order.  The parameters of the fiducial
cosmological simulation are those of the B2 model in
\cite{vielhaehnelt06}. As before the flux statistics have been
corrected for box-size and resolution effects.  We compute the
derivatives required for the Taylor expansion by performing between
four and six hydrodynamical simulations for every cosmological and
astrophysical parameter considered.  In addition, we now allow for the
effect of the reionization redshift, $z_{\rm re}$, and introduce this
as an extra parameter. We interpolate between two very different
reionization histories with $z_{\rm re} \sim 15$ and $z_{\rm re} \sim
7$. We also introduce two extra parameters describing the redshift
evolution of the thermal state of the IGM, the power--law index of the
$T$ and $\gamma$ relations at $z>3$ (a redshift range which is not
probed by the PDF).

\begin{table}
\begin{tabular}{llll}
\hline
\small{Param.} & \tiny{B2$_1^{(lowz)}$} & \tiny{B2$_1^{(allz)}$} & \tiny{B2$_1^{(lowz)}$+(20,256)$^{0.1-0.8}$}\\
\hline
$\sigma_8$ & $0.85 \pm 0.05$  & $0.86 \pm 0.04$ & $0.90\pm0.02$ \\
$n_{\rm s}$ & $0.93 \pm 0.03$ & $0.96\pm 0.02$ & $0.95\pm 0.02$ \\
$\Omega_{\rm 0m}$ & $0.25 \pm 0.04$ & $0.26 \pm 0.03$ & $0.25\pm0.03$ \\
$H_0$ & $78 \pm 7$ & $ 77\pm 7$  & $80\pm 5$ \\
$T_0$ & $38 \pm 7$ & $ 42 \pm 6$ & $26 \pm 4$\\
$T_0^{s1}$ & $-0.6 \pm 1.3$ & $ -0.3 \pm 1.1 $ & $1.4\pm 0.5$ \\
$T_0^{s2}$ & $-2.3 \pm 1.3$ & $-3.9 \pm 1.3 $ & $-3.1\pm 1.5$\\
$\gamma^A$ & $0.63 \pm 0.50$ & $ 0.79 \pm 0.51$ & $0.50 \pm 0. 20$ \\
$\gamma^{s1}$ & $-0.7 \pm 2.6$ & $0.4 \pm 2$ &  $-2.1\pm 1.6$\\
$\gamma^{s2}$ & $-1.2 \pm 2.1$ & $-1.4 \pm 1.5$ & $-1.4\pm 1.6$ \\
$\tau_{\rm eff}^A$  & $0.326 \pm 0.028$ & $0.322 \pm 0.028$ & $0.320 \pm0.007$ \\
$\tau_{\rm eff}^{s1}$ & $3.19 \pm 0.25$ & $3.25 \pm 0.23$ & $3.12 \pm 0.10$\\
$z_{\rm re}$  & $11.9 \pm  3.8$ & $9.1 \pm 2.7$ & $14.3 \pm 3.6$\\
\hline
$\chi^2/d.o.f.$ &  78.9/85 & 139.7/121 & 136/130\\
\end{tabular}
\caption{Cosmological and astrophysical parameters derived from the
  B2$_1$ model ($\gamma \sim 1$) for the flux power spectrum: $(lowz)$ flux PS
  fitted in the range $z=[2.2-3.6]$; $(allz)$ flux PS fitted in the
  range $z=[2.2-4.2]$. The probability of having a $\chi^2$ value
  larger than this for model B2$_1^{(lowz)}$ is 64 \%, while for model
  B2$_1^{(allz)}$ it is 12 \%. The constraints for a joint analysis of
  flux PS and PDF ($F=0.1-0.8$) are shown in the 3$^{rd}$ column (prob. is 35\%).\label{tab3}}
\end{table}

The results for the power spectrum only analysis are summarised in the
first two columns of Table \ref{tab3} for a low redshift only sample,
$z=[2.2-3.6]$, and the full SDSS data set, $z=[2.2-4.2]$. We decided
to perform a separate analysis which omits the highest redshift bins
following \cite{vielhaehnelt06}, who obtained a somewhat poorer fit
for the high redshift ($z>3.6$) PS estimates.  Despite the fact that
the flux statistics were calculated by expanding around a reference
model with very different thermal history, in both instances the
analysis still gives constraints on the cosmological parameters that
are in agreement with the previous analysis of
\cite{vielhaehnelt06}.  This is rather reassuring. However, there are
some aspects of the results that need scrutiny.  First, we note that
for the flux power only the temperature at mean density, $T_0$, is
significantly higher than that preferred by the PDF (and higher than
expected for the photoionized IGM).  Secondly, the value of $\sigma_8$
is now somewhat on the lower end of values allowed by the previous analysis of
\lya data and thus in better agreement with the CMB data
(e.g. \citealt{komatsu08}).  This is due to the degeneracy between
$\sigma_8$ and $\gamma$ discussed in B08; allowing for $\gamma<1$
means that the flux power can now be fitted by a slightly lower
$\sigma_8$ (but note that other parameters also have a a significant
influence on the inferred $\sigma_8$, most notably the mean flux
level).

Overall the results from the new flux PS analysis are consistent with
those inferred from the PDF alone except for the value of $T_0$ at
$z=3$, which is several $\sigma$ above that inferred from the best fit
to the flux PDF.  In the last column of Table \ref{tab3} we show the
constraints for a joint analysis of flux PDF and PS. Somewhat
surprisingly the joint analysis prefers a larger value of
$\sigma_8=0.9\pm 0.02$ with rather small errors. We explicitly checked
that this large value of $\sigma_8$ is related to the rather different
$T_0^A$ values that the PDF and PS favour.  If we artificially remove
the constraint of the temperature being simultaneously consistent
with the somewhat discrepant temperatures favoured by the flux PDF and
PS, the joint analysis gives $\sigma_8=0.86\pm 0.03$ with an
improvement of $\Delta \chi^2 = 12$.  A not yet accounted for
systematic error in the measurement of the flux PDF and/or PS appears
to be a possible explanation for this discrepancy.  Alternatively the
inconsistencies may suggest that a power--law $T-\rho$ relation is a
poor approximation to the thermal state of the IGM and a wider range
of physically motivated relations should be considered in future
simulations.

\section{Discussion}

We have presented cosmological and astrophysical constraints derived
from the K07 \lya flux PDF measured from a set of 18 high-resolution
QSO spectra whose statistical and systematic errors have been
carefully estimated.  The \lya flux PDF on its own provides tight
constraints on the thermal state of the IGM and on cosmological
parameters describing the linear dark matter PS. The results have been
obtained by fitting the flux PDF at three different redshift bins in
the range $2<z<3$ and for three different flux ranges $F=[0.1-0.8]$,
$F=[0-0.9]$, and $F=[0-1]$. There is good agreement between the
analyses for the full flux range and the two restricted flux ranges
and the results are consistent with those derived in the simpler
analysis made by B08.  An inverted temperature-density relation is
favoured at the $\sim 3\sigma$ level (at $z \sim 3$) if we consider the
PDF for the full flux range, but the significance is reduced to
$1-1.5\sigma$ if we restrict the analysis to $F=[0.1-0.8]$.  The
constraints for other parameters are in agreement with those presented
in the literature ({\it e.g.}  the SDSS flux power in
\citealt{mcdonald05}). We have also refined the method used by
\cite{vielhaehnelt06} and updated the constraints from the SDSS flux
power spectrum.  The constraints from the flux power spectrum are
consistent with those from the flux PDF, with the exception that the
flux power spectrum prefers a significantly larger temperature at mean
density, $T_0$.  A joint PS--PDF analysis gives a reasonable fit to
the data but results in a larger $\sigma_8$ than an analysis of CMB
data alone. This discrepancy appears to be related to the higher $T_0$
that the flux power spectrum prefers.

Recent simulations of photo-heating during \HeII reionization indicate
that an inverted $T-\rho$ relation is very difficult to achieve
within the standard model describing the thermal state of the IGM 
even if radiative transfer effects are taken into account, 
at least if \HeII reionization is driven primarily by QSOs
(\citealt{mcquinn09,Bolton09b}).  It therefore appears difficult to
reproduce the observed flux PDF without invoking a not yet identified  source of
IGM heating, or additional systematic errors which impact on the flux 
PDF.  Similarly, the high $T_0$ values preferred by the PS
are very difficult to reconcile with constraints from the widths of
thermally broadened absorption lines (e.g. \cite{schaye00,ricotti00})
and our understanding of the heating of the photoionized IGM.

As the results from the flux PDF
and the CMB data agree very well, the rather high values of $T_0$ 
preferred by the flux power spectrum suggests perhaps instead that we have   
identified a not yet accounted for systematic  error in 
the SDSS flux power spectrum data.
Independent analysis based on line statistics, on new data sets at
medium and high--resolution and further progress in incorporating \HeII
reionization models into high resolution hydrodynamical simulations
will hopefully allow us to further improve our  understanding of  the systematic
uncertainties of \lya forest data and resolve these small but
statistically significant inconsistencies.

\section*{Acknowledgments.}

Numerical computations were performed on the COSMOS supercomputer at
DAMTP and on the High Performance Computer Cluster Darwin (HPCS) in
Cambridge (UK). COSMOS is a UK-CCC facility which is supported by
HEFCE, PPARC and Silicon Graphics/Cray Research. Part of the analysis
was also performed at CINECA (Italy) with CPU time assigned thanks to
an INAF-CINECA grant. We thank A. Lidz and P. McDonald for suggestions
and useful criticism.

\bibliographystyle{mn2e}
 \bibliography{master2.bib}

\end{document}